\documentclass[final,5p,times,twocolumn]{elsarticle}

\usepackage{amssymb}
\usepackage{amsmath}

\journal{Nuclear Physics B}

\begin{document}

\begin{frontmatter}



\title{Surrogate neutron-capture studies with fission detection in inverse kinematics at the experimental storage ring ESR}

\author[LP2IB]{Bogusław Włoch}
\author[LP2IB]{Camille Berthelot}
\author[LP2IB]{Guy Leckenby}
\author[LP2IB]{Beatriz Jurado}
\author[LP2IB]{Jerome Pibernat}
\author[MPIK]{Manfred Grieser}
\author[GSI]{Jan Glorius}
\author[GSI,Koln]{Yury A. Litvinov}
\author[IJCLab]{Laurent Audouin}
\author[LP2IB]{Bertram Blank}
\author[LP2IB]{Lucas Bégué--Guillou}
\author[GANIL,Caen]{Alex Cobo Zarzuelo}
\author[JWGU]{Sophia Florence Dellmann}
\author[CEA-DAM,Bruyeres]{Marc Dupuis}
\author[Jena]{Oliver Forstner}
\author[GANIL,Caen]{Alexis Francheteau}
\author[Koln]{David Freire Fernández}
\author[Osaka]{Miki Fukutome}
\author[LP2IB]{Mathias Gerbaux}
\author[LP2IB]{Jérôme Giovinazzo}
\author[GSI]{Alexandre Gumberidze}
\author[Swe]{Andreas Heinz}
\author[FRIB]{Ana Henriques}
\author[GSI]{Regina Hess}
\author[GANIL,Caen]{Indu Jangid}
\author[GSI]{Anton Kalinin}
\author[Saclay]{Wolfram Korten}
\author[GSI]{Sergey Litvinov}
\author[GSI]{Bernd Lorentz}
\author[Sevilla]{Antonio M. Moro}
\author[GSI]{Nikolaos Petridis}
\author[GSI]{Ulrich Popp}
\author[Sevilla]{Gregory Potel}
\author[GANIL]{Diego Ramos}
\author[LP2IB]{Mathieu Roche}
\author[GSI]{Mohammad Shahab Sanjari}
\author[IJCLab]{Michele Sguazzin}
\author[Surrey]{Ragandeep Singh Sidhu}
\author[GSI]{Uwe Spillmann}
\author[GSI]{Markus Steck\footnote{Deceased}}
\author[Jena]{Thomas Stöhlker}
\author[Saitama]{Takayuki Yamaguchi}

\affiliation[LP2IB]{organization={Université de Bordeaux, CNRS, LP2I Bordeaux},
            city={Gradignan},
            postcode={33170}, 
            country={France}}

\affiliation[MPIK]{organization={Max-Planck-Institut fur Kernphysik},
            city={Heidelberg},
            postcode={69117}, 
            country={Germany}}

\affiliation[GSI]{organization={GSI Helmholtzzentrum fur Schwerionenforschung},
            city={Darmstadt},
            postcode={64291}, 
            country={Germany}}

 \affiliation[Saclay]{organization={CEA Paris-Saclay - DRF/IRFU/DPhN},
            city={Gif-sur-Yvette},
            postcode={91191}, 
            country={France}}

\affiliation[GANIL]{organization={GANIL, CRNS/IN2P3-CEA/DRF},
            city={Caen},
            postcode={14000}, 
            country={France}}

\affiliation[Caen]{organization={Université de Caen-Normandie},
            city={Caen},
            postcode={14000}, 
            country={France}}

\affiliation[JWGU]{organization={Goethe Universitat Frankfurt},
            city={Frankfurt am Main},
            postcode={60438}, 
            country={Germany}}

\affiliation[CEA-DAM]{organization={Commissariat à l'énergie atomique et aux énergies alternatives, DAM, DIF},
            city={Arpajon},
            postcode={91297}, 
            country={France}}

\affiliation[Bruyeres]{organization={Université Paris-Saclay, LMCE},
            city={Bruyères-Le-Châtel},
            postcode={91680}, 
            country={France}}

\affiliation[Jena]{organization={Friedrich-Schiller-Universität Jena},
            city={Jena},
            postcode={07743}, 
            country={Germany}}

\affiliation[Koln]{organization={Universität zu Köln},
            city={Cologne},
            postcode={50923}, 
            country={Germany}}
            
\affiliation[Osaka]{organization={Osaka University},
            city={Osaka},
            postcode={565-0871}, 
            country={Japan}}

\affiliation[IJCLab]{organization={Université Paris-Saclay, CNRS, IJCLab},
            city={Orsay},
            postcode={91405}, 
            country={France}}




\affiliation[Swe]{organization={Chalmers University of Technology},
            city={Gothenburg},
            postcode={41296}, 
            country={Sweden}}
            
\affiliation[FRIB]{organization={Facility for Rare Isotope Beams, Michigan State University},
            city={East Lansing},
            postcode={48824-1321}, 
            country={United States}}

\affiliation[Aachen]{organization={FH Aachen University of Applied Sciences},
            city={Aachen},
            postcode={52066}, 
            country={Germany}}

\affiliation[Sevilla]{organization={Universidad de Sevilla},
            city={Sevilla},
            postcode={41080}, 
            country={Spain}}

\affiliation[Surrey]{organization={University of Surrey},
            city={Guildford},
            postcode={GU2 7XH}, 
            country={United Kingdom}}
            
\affiliation[Saitama]{organization={Saitama University},
            city={Saitama},
            postcode={338-8570}, 
            country={Japan}}

\begin{abstract}

The NECTAR (Nuclear rEaCTions At storage Rings) experiment at the ESR heavy-ion storage ring at GSI/FAIR Darmstadt is dedicated to surrogate reaction studies of neutron-induced reactions on heavy nuclei in inverse kinematics. 
In this work, we report on the implementation and performance of a newly developed fission-fragment detection system integrated into the NECTAR experimental setup. The upgraded detector configuration enables, for the first time in a surrogate experiment, the simultaneous detection of \(\gamma\)-decay residues, multi-neutron-emission residues, and fission fragments. The full setup was used for the first time in an experiment where a stored beam of bare \(^{238}\)U\(^{92+}\) ions at 17.24~MeV/u interacted with a gas-jet deuterium target, populating excited \(^{238}\)U and \(^{239}\)U nuclei via the \(^{238}\)U(d,d') and \(^{238}\)U(d,p) reactions. We describe the geometry of the used fission fragment detectors, design constraints, and simulation-based efficiency determination. The target-like particle identification and beam-like residue spectra demonstrating the performance of the complete setup are also shown.

\end{abstract}


\end{frontmatter}

\section{Introduction}
\label{sec_Intro}

Surrogate reaction techniques provide a powerful alternative for determining neutron-induced reaction cross sections of nuclei for which direct measurements are impractical or, at least at present, impossible, in particular short-lived or highly radioactive nuclei. 
In these methods, the same compound nucleus as formed in a neutron-induced reaction is populated through an alternative reaction mechanism, and its decay probabilities are measured as a function of excitation energy. 
By determining these decay probabilities, fundamental nuclear properties such as the nuclear level density (NLD), fission barrier heights, and the \(\gamma\)-strength function (\(\gamma\)SF) can be constrained~\cite{Surr1, Surr2}. 
These quantities serve as critical inputs to reaction models and enable more reliable calculations of neutron-induced cross sections. 
Surrogate reactions are typically performed in direct kinematics.
Decay probabilities are inferred from the detection of emitted  \(\gamma\) rays or fission fragments using detector arrays surrounding the target~\cite{Sanchez, Ducasse}. Although these approaches have been successfully applied, they often require substantial efficiency corrections and background subtraction.

The NECTAR (Nuclear rEaCTions At storage Rings)~\cite{NECTAR} experimental program exploits the unique capabilities of heavy-ion storage rings to perform surrogate reaction measurements in inverse kinematics.  The first NECTAR experiments, conducted in 2022, demonstrated the feasibility of this approach at the ESR experimental storage ring at GSI/FAIR Darmstadt,~\cite{ESR} enabling for the first time measurements of \(\gamma\)-decay and neutron-emission probabilities following surrogate reactions on \(^{208}\)Pb~\cite{Nectar_PRL, Nectar_PRC}. 
A key limitation of the 2022 setup was the absence of direct fission-fragment detection, which would prevent a complete determination of decay probabilities for fissionable systems. 
The present work addresses this limitation by introducing a dedicated fission-fragment detection system. 
The primary focus of this article is the design, implementation, and validation of this new detector system and its integration into the existing NECTAR setup.

\section{Experimental setup}
\label{sec_Exp}

\begin{figure}[h]
\centerline{\includegraphics[width=.47 \textwidth]{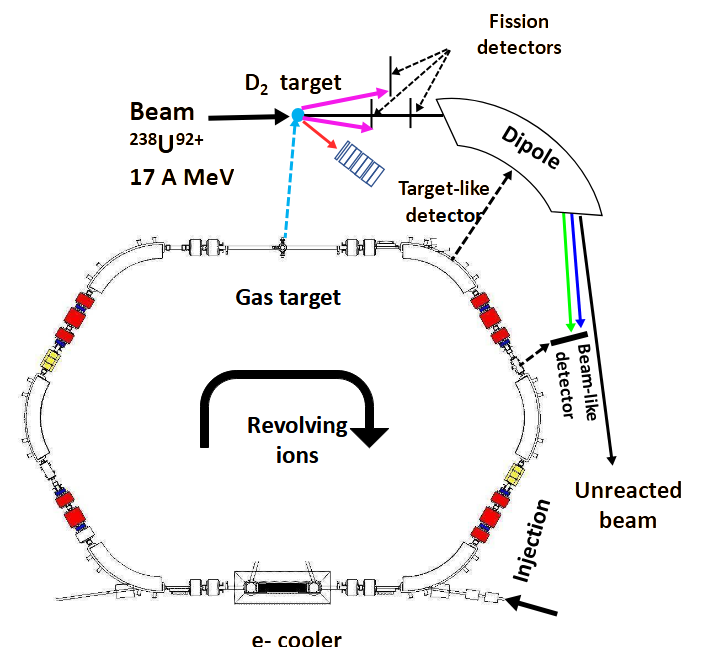}}
\caption{Schematic overview of the ESR and the NECTAR experimental setup. The red, pink, blue and green arrows represent the trajectories of the target-like residues, fission fragments and heavy residues formed after \(\gamma\) and neutron emission, respectively.}
\label{schem_NECTAR}
\end{figure}

The experimental setup combined three complementary detector systems: a light-particle telescope for reaction-channel identification, a newly developed fission-fragment detection system downstream of the target, and a beam-like residue detector located in a dispersive section of the ESR~\cite{Rings}. 
To avoid degradation of the ultra-high vacuum,
all the detectors were housed in pockets behind 25~\(\mu\)m thick stainless-steel windows, through which the reaction products could pass. 
A schematic overview of the experimental setup is presented in Figure~\ref{schem_NECTAR}.

The full setup was used for the first time at the ESR using a stored beam of bare \(^{238}\)U\(^{92+}\) ions with an energy of 17.24~MeV/u. 
The beam repeatedly interacted with an internal deuterium gas-jet target, populating excited compound nuclei via the reactions
\begin{align}
^{238}\mathrm{U}(d,p)^{239}\mathrm{U}^{*}, \\
^{238}\mathrm{U}(d,d')^{238}\mathrm{U}^{*}.
\end{align}

\subsection{Target-like particle telescope}

Target-like reaction products were detected using a dedicated telescope placed in the horizontal plane of the ring at a laboratory angle of \(\theta = 60^\circ\). 
The telescope provided event-by-event particle identification and excitation-energy reconstruction, and its design followed the configuration successfully employed in the 2022 NECTAR experiment~\cite{Nectar_PRC}.

The telescope consisted of a 306~\(\mu\)m thick BB8 double-sided silicon strip detector (DSSSD) from Micron Semiconductor Ltd., segmented into 16 vertical and 16 horizontal strips over an active area of \(20 \times 20~\mathrm{mm}^2\)~\cite{BB8}. It operated as a \(\Delta E\) detector.  It was followed by a stack of six single-area silicon detectors $E_i$ (MSX04 model), each with a thickness of 1.5~mm~\cite{MSX04}.

\begin{figure}[h]
\centerline{\includegraphics[width=.47 \textwidth]{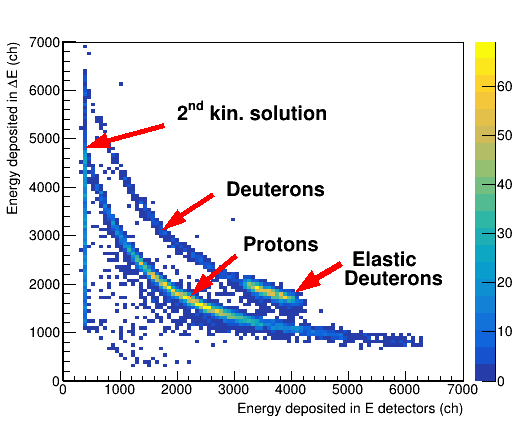}}
\caption{\(\Delta E\)--\(E\) spectrum from the target-like particle telescope (vertical strip 13), demonstrating separation of protons and deuterons.}
\label{telescope_id}
\end{figure}

Figure~\ref{telescope_id} shows a standard identification  \(\Delta E\)--\(E\) spectrum, where  \(\Delta E\) is the signal from the thin BB8 detector and $E$ is the sum of the signals from $E_1 ... E_6$ detectors. Only events that hit vertical strip number 13 are shown. Proton and deuteron bands are clearly separated, allowing a clear selection of the \(^{238}\)U(d,p) and \(^{238}\)U(d,d') reaction channels.
For higher energy particles that can punch through the first \(E\) detector, we can additionally improve the identification by making similar correlations using other pairs of detectors, e.g., \(E_1\) vs. \(E_2\), etc.

\subsection{Fission-fragment detection system}
\label{subsec:fission}

The central upgrade of the NECTAR setup is the implementation of a dedicated fission-fragment detection system downstream of the gas-jet target. 
The design of this system was constrained by the limited available space, the ultra-high vacuum requirements of the ESR, and the need to protect detectors during beam injection.

Three fission detectors were installed in dedicated pockets mounted in fittings of the target chamber. Two identical detectors were positioned above and below the beam axis, and a third detector was mounted on the horizontal plane on the outside of the ring. 
Together, these detectors provide large solid-angle coverage for forward-focused fission fragments.

\begin{figure}[h]
\centerline{\includegraphics[width=.47 \textwidth]{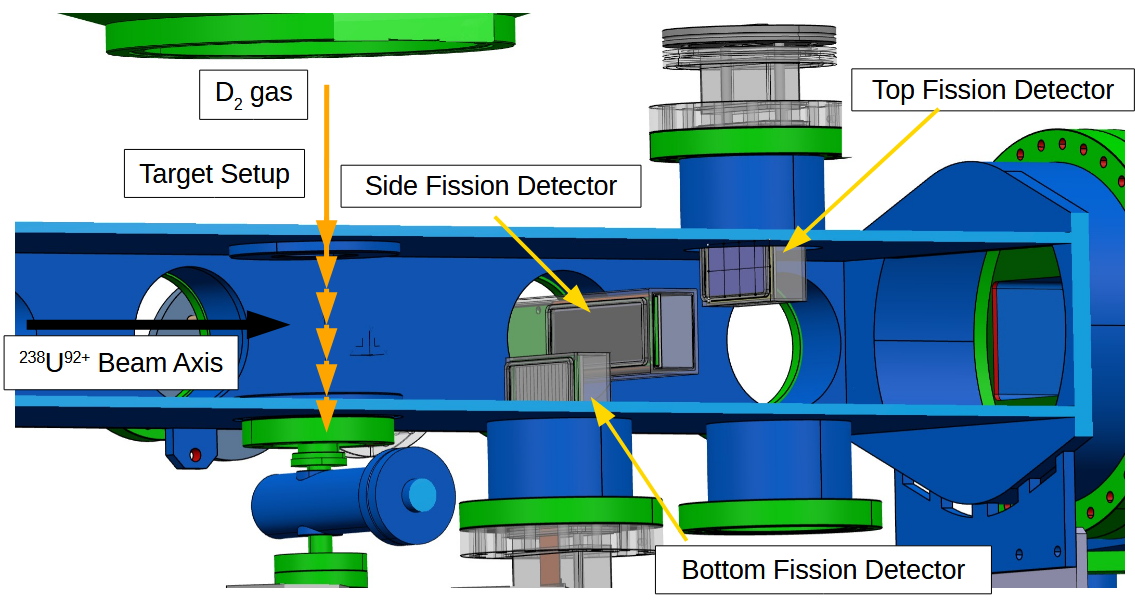}}
\caption{Cross section of the 3D model of the experimental chamber showing the integration of the fission detectors. The target-like detector (not visible) would be on the side of the viewer.}
\label{schem_chamber}
\end{figure}

The bottom, top, and side pockets were located at distances of 220~mm, 400~mm, and 340~mm from the target center, respectively. Due to geometrical constraints imposed by the chamber fittings, the side detector required a distinct mechanical design. A three-dimensional visualization of the chamber and pocket arrangement is shown in Fig.~\ref{schem_chamber}.

To prevent detector damage during beam injection, the side fission pocket was mounted on a movable system allowing it to be fully retracted during injection and inserted close to the beam axis only during data taking. 
The motion was achieved using a bellow system that preserved ultra-high vacuum conditions.


The top and bottom fission detectors were custom-designed BB36 DSSSDs from Micron Semiconductor Ltd., each with an active area of \(80 \times 40~\mathrm{mm}^2\) with a thickness of 500~\(\mu\)m and a segmentation of 16 vertical by 16 horizontal strips. They were operated at a bias voltage of 100~V.

The side fission detector was a BB29 DSSSD with a thickness of 500~\(\mu\)m and an active area of \(122 \times 40~\mathrm{mm}^2\), segmented into 122 vertical and 40 horizontal strips. This detector design had previously been employed as the beam-like residue detector in the 2022 NECTAR experiment, facilitating reuse of existing electronics and mechanical components.


The fission detection system was studied using Monte Carlo simulations based on G4beamline~\cite{G4beamline} and the GEF code \cite{GEF}. 
While fission fragments are emitted in all directions in the center-of-mass frame, the high beam velocity leads to strong forward focusing in the laboratory frame. 
For both \(^{238}\)U and \(^{239}\)U fissioning systems, the fragments are confined within a cone of approximately 17$^\circ$ around the beam axis.

The kinetic energies of the fission fragments in the laboratory frame vary between 9~MeV/u and 29~MeV/u, and practically do not depend on the excitation energy of the nucleus. 
At these energies, all the fragments went through the stainless-steel pocket windows and could reach the detectors with sufficient energy to ensure 100\% intrinsic detection efficiency.

\begin{figure}[h]

\centerline{\includegraphics[width=.40 \textwidth]{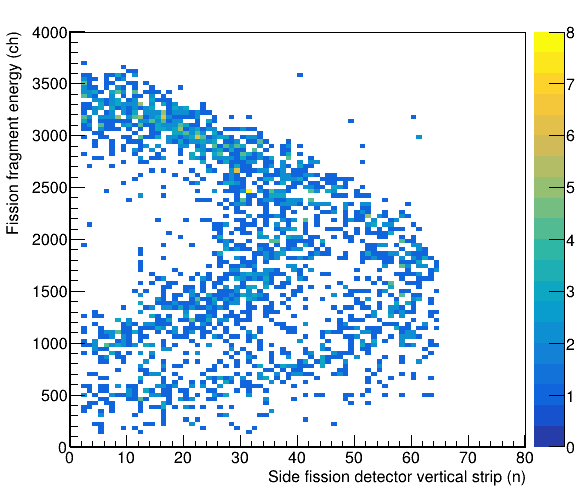}}
\centerline{\includegraphics[width=.40 \textwidth]{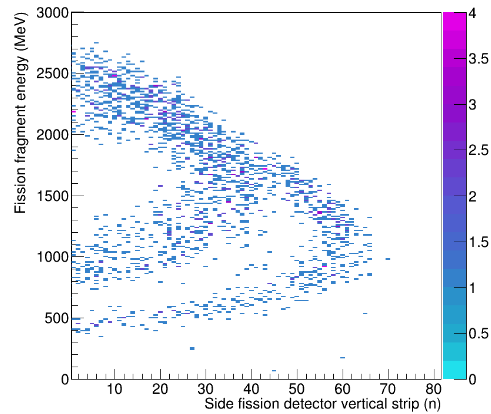}}
\caption{Comparison of measured (top) and simulated (bottom) fission-fragment hit distributions on the side detector. The energy of the fragments is represented as a function of their horizontal position measured with the vertical strips of the detector.}
\label{Fission_comparison}
\end{figure}

The full geometry of the reaction chamber, fission pockets, and detectors was implemented in the simulations. 
Figure~\ref{Fission_comparison} compares simulated and measured amplitude and hit distributions on the side fission detector. 
The clearly visible two-ring pattern comes from the detection of a light and a heavy fission fragment originating from the asymmetric fission mode of \(^{239}\)U, showing excellent agreement and validating the implemented geometry.
The resulting fission detection efficiency was determined to be about 64\% .

\subsection{Beam-like residue detector}

Different beam-like residues were separated using the ESR dipole magnet based on the difference in their magnetic rigidity and detected in a straight section of the ring. 
The detector employed the same BB29 DSSSD model design as the side fission detector, with minor differences in the mechanical supports.

The intrinsic efficiency of the detector was 100\% for all residues reaching the detection plane. 
For the \(^{238}\)U(d,p) reaction, the Monte Carlo simulations showed full transmission of beam-like residues for \(\gamma\)-emission, one-neutron-emission and two-neutron-emission. 
This implies an efficiency for determining \(\gamma\), one-neutron and two-neutron emission probabilities of 100\%, demonstrating the unique capabilities of NECTAR. 
The transmission for three-neutron-emission was below 100\% because some of the heavy residues' trajectories hit the wall of the ring beam pipe. The beam-like residues produced after four-neutron-emission were deflected too much inside the ring and none of them was able to reach the beam-like residue detector.

\begin{figure}[!h]
\centerline{\includegraphics[width=.47 \textwidth]{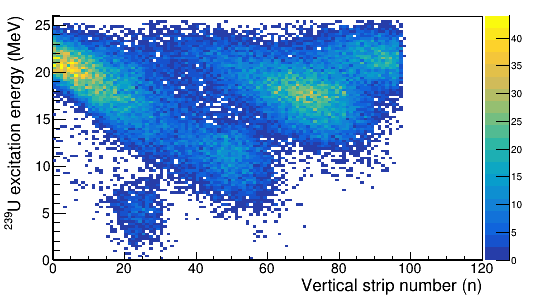}}
\caption{Beam-like detector identification spectra. The horizontal axis represents the horizontal position of beam-like residues after \(^{238}\)U(d,p) reaction. The vertical axis is the excitation energy of the compound nucleus.}
\label{HR_dp_all}
\centerline{\includegraphics[width=.47 \textwidth]{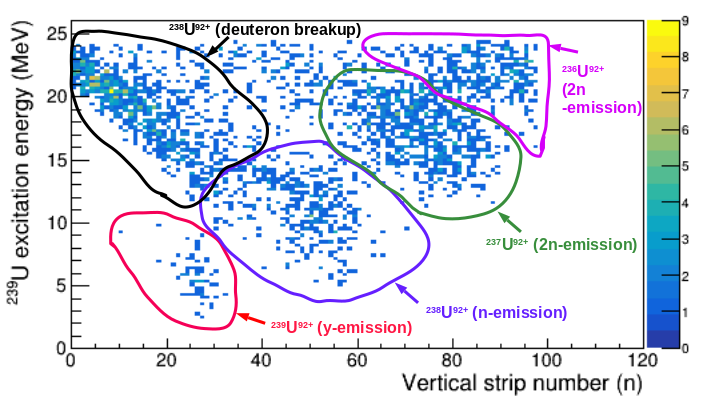}}
\caption{Similar as in Figure \ref{HR_dp_all} but in coincidence with a
proton detected in the 6th vertical strip of the target-like telescope detector. The marked areas indicate the different uranium isotopes and thus the different decay channels of the compound nucleus.}
\label{HR_dp}
\end{figure} 

\section{Beam-like residue identification spectra}

The different decay channels of the compound nucleus formed in the \(^{238}\)U(d,p) reaction can be identified when plotting the vertical strip (horizontal position) of a hit in the beam-like detector versus the excitation energy of the compound nucleus (Figure~\ref{HR_dp_all}). 
The strips are numbered sequentially from zero to 122, with zero denoting the strip closest to the beam axis. 

Different uranium isotopes form distinct bands that move closer to the edge of the detector as the excitation energy of the nucleus increases. 
The structures corresponding to the different decays are clearly visible, but there is also a relatively large amount of mixing between them. 
Therefore, to improve separation, we decided to repeat this procedure separately for each vertical strip of the target-like telescope. 
This narrows the angular width of the detected target-like particles, thereby also reducing the spread of the peaks detected in the beam-like detector, as shown in Figure~\ref{HR_dp}.

Below approximately 5~MeV of excitation energy, only \(\gamma\)-decay residues are observed. Above the neutron separation energy, distinct structures corresponding to one-, two-, and three-neutron emission appear at the expected thresholds.

At the highest excitation energies, a pronounced structure associated with heavy residues produced after deuteron breakup is observed.  It is located closer to the beam axis, at strip numbers smaller than 60. 
Breakup events are those in which, upon collision with a uranium nucleus, a deuteron splits into a proton and a neutron, leaving the nucleus without excitation.
The ability to identify events originating from the deuteron breakup is a truly unique feature of the NECTAR experiment, as this has been impossible in all previous surrogate experiments with deuteron target in direct kinematics. 
Further analysis of this experiment will allow these events to be used to subtract the breakup contribution from the compound nucleus decay probabilities.

\section{Conclusions}

A dedicated fission-fragment detection system has been successfully designed, implemented, and commissioned for the NECTAR experiment at the ESR. 
The upgraded setup enables, for the first time, the simultaneous measurement of \(\gamma\)-decay, multi-neutron-emission, and fission channels in a storage ring based surrogate reaction experiment.
The detector performance and efficiency have been validated through Monte Carlo simulations and comparison with experimental data.
Target-like and beam-like residue identification spectra demonstrated the ability to identify all available decay channels.
This development represents a crucial step toward complete surrogate measurements of neutron-induced reactions on heavy nuclei in inverse kinematics.
The final goal of the NECTAR project is to apply this newly established
technique to radioactive beams. The storage of such secondary beams in
the ESR at intensities needed for nuclear reaction studies has already been
demonstrated~\cite{Jan,Dellmann}. 
As a final step the angular coverage for the target-like detection will be increased by using larger and more numerous particle telescopes.
The surrogate reaction measurements are an essential part of a broader research programme at ion storage ring of the ESR and CRYRING \cite{Letinsky} aiming at nuclear reaction studies \cite{Jan,Dellmann,carme}.

\section*{Acknowledgments}


This work is supported by the European Research Council (ERC) under the European Union’s Horizon 2020 research and innovation programme (ERC-Advanced grant NECTAR, grant agreement No 884715). The results presented here are based on the experiment G-22-00028, which was performed at the ESR storage ring of the GSI Helmholtzzentrum für Schwerionenforschung, Darmstadt (Germany) in the context of FAIR Phase-0. We warmly thank fot the financial support from the GSI/IN2P3 collaboration 19-80. This project has also received funding from the European Union’s Horizon Europe Research and Innovation programme under Grant Agreement No 101057511 (EURO-LABS).
AH is grateful for funding from the Knut and Alice Wallenberg Fundation under KAW 2020.0076.


\begin{thebibliography}{00}

\bibitem{Surr1} J. D. Cramer and H. C. Britt, Nucl. Sci. Eng. 41, 177 (1970).
\bibitem{Surr2} J. E. Escher et al., Rev. Mod. Phys. 84, 353 (2012).
\bibitem{Sanchez} R. Perez Sanchez et al., Phys. Rev. Lett. 125, 122502 (2020).
\bibitem{Ducasse} Q. Ducasse et al., Phys. Rev. C 94, 024614 (2016).
\bibitem{NECTAR} https://www.lp2ib.in2p3.fr/nucleaire/nex/erc-nectar/
\bibitem{ESR} B. Franzke, Nucl. Instrum. Methods B 25, 18 (1987).
\bibitem{Nectar_PRL} M. Sguazzin et al., Phys. Rev. Lett. 134, 072501 (2025).
\bibitem{Nectar_PRC} M. Sguazzin et al., Phys. Rev. C 111, 024614 (2025).
\bibitem{Rings} Markus Steck, Yuri A. Litvinov, Prog. Part. Nucl. Phys., 115, 103811, (2020). 

\bibitem{BB8} https://www.micronsemiconductor.co.uk/product/bb8/

\bibitem{MSX04} https://www.micronsemiconductor.co.uk/product/msx04/



\bibitem{G4beamline} G4beamline – http://g4beamline.muonsinc.com
  \bibitem{GEF} K.-H. Schmidt B. Jurado, Ch. Amouroux, C. Schmitt, Nuclear Data Sheets 131 (2016).

\bibitem{Jan} J. Glorius et al., Nucl. Instrum. Methods Phys. Res. B, 541, (2023).
\bibitem{Dellmann} Dellmann et al., Phys. Rev. Lett., 134, 142701 (2025).
\bibitem{Letinsky} Lestinsky, M., Andrianov, V., Aurand, B. et al. Physics book: CRYRING@ESR. Eur. Phys. J. Spec. Top. 225, 797–882 (2016).
\bibitem{carme} C.G. Bruno et al., Nucl. Instrum. Methods Phys. Res. A, 1048, (2023).

\end{thebibliography}
\end{document}